\documentclass[fleqn,usenatbib]{mnras}

\usepackage{newtxtext,newtxmath}
\usepackage[T1]{fontenc}
\usepackage{ae,aecompl}
\usepackage{subfig}

\usepackage{graphicx}	% Including figure files
\usepackage{amsmath}	% Advanced maths commands
\usepackage{amssymb}	% Extra maths symbols

\usepackage{multicol}

\newcommand{\XMM}{ XMM-{\em Newton}}
\newcommand{\Chandra}{{\em Chandra}}

\title[Dramatic spectral variability of a QSO at $z\sim 1$]{Dramatic X-ray spectral variability of a Compton-thick type-1 QSO at $z\sim 1$}

\author[T. Simm et al.]{
T. Simm$^{1}$\thanks{E-mail: tsimm@mpe.mpg.de},
J. Buchner$^{2,3}$,
A. Merloni$^{1}$,
K. Nandra$^{1}$,
Y. Shen$^{4}$,
T. Erben$^{5}$,
A. L. Coil$^{6}$,
\newauthor
C. N. A. Willmer$^{7}$,
and D. P. Schneider$^{8,9}$
\\
$^{1}$Max Planck Institute for Extraterrestrial Physics, Giessenbachstrasse, Postfach 1312, 85741 Garching, Germany\\
$^{2}$Pontificia Universidad Catolica de Chile, Instituto de Astrofisica, Casilla 306, Santiago 22, Chile\\
$^{3}$Excellence Cluster Universe, Boltzmannstr. 2, D-85748, Garching, Germany\\
$^{4}$Department of Astronomy, University of Illinois at Urbana-Champaign, 1002 W Green Street, Urbana, IL 61801\\
$^{5}$Argelander-Institut f\"ur Astronomie, Auf dem H\"ugel 71, D-53121 Bonn, Germany\\ 
$^{6}$Center for Astrophysics and Space Sciences, Dpt. of Physics, Univ. of California San Diego, 9500 Gilman Drive, La Jolla, CA 92093-0424\\
$^{7}$Steward Observatory, University of Arizona, 933 N. Cherry Avenue, Tucson, AZ 85721, USA\\ 
$^{8}$Department of Astronomy and Astrophysics, The Pennsylvania State University, University Park, PA 16802\\
$^{9}$Institute for Gravitation and the Cosmos, The Pennsylvania State University, University Park, PA 16802\\
}

\date{Accepted 10.08.2018. Received 08.08.2018; in original form 12.03.2018}

\pubyear{2018}

\hypersetup{draft}

\begin{document}
\label{firstpage}
\pagerange{\pageref{firstpage}--\pageref{lastpage}}
\maketitle

% Abstract of the paper
\begin{abstract}
We report on the discovery of a dramatic X-ray spectral variability event observed in a $z\sim 1$ broad line type-1 QSO. The \XMM\, spectrum from the year 2000 is characterized by an unobscured power-law spectrum with photon index of $\Gamma\sim 2$, a column density of $N_{\mathrm{H}}\sim 5\times 10^{20}\,\mathrm{cm^{-2}}$, and no prominent reflection component. Five years later, \Chandra\, captured the source in a heavily-obscured, reflection-dominated state. The observed X-ray spectral variability could be caused by a Compton-thick cloud with $N_{\mathrm{H}}\sim 2\times 10^{24}\,\mathrm{cm^{-2}}$ eclipsing the direct emission of the hot corona, implying an extreme $N_{\mathrm{H}}$ variation never before observed in a type-1 QSO. An alternative scenario is a corona that switched off in between the observations. In addition, both explanations require a significant change of the X-ray luminosity prior to the obscuration or fading of the corona and/or a change of the relative geometry of the source/reflector system. Dramatic X-ray spectral variability of this kind could be quite common in type-1 QSOs, considering the relatively few datasets in which such an event could have been identified. Our analysis implies that there may be a population of type-1 QSOs which are Compton-thick in the X-rays when observed at any given time. 
\end{abstract}

% Select between one and six entries from the list of approved keywords.
% Don't make up new ones.

%%%%%%%%%%%%%%%%%%%%%%%%%%%%%%%%%%%%%%%%%%%%%%%%%%

%%%%%%%%%%%%%%%%% BODY OF PAPER %%%%%%%%%%%%%%%%%%

\section{Introduction}

Over the last two decades evidence has emerged for variations of the line of sight (LOS) column density, ($N_{\mathrm{H}}$), in nearby type-2 (obscured) active galactic nuclei (AGN). These are inferred from the shape of the X-ray spectrum, which changes dramatically if a source transitions from a transmission-dominated Compton-thin ($N_{\mathrm{H}}<10^{24}\,\mathrm{cm^{-2}}$) state to a reflection-dominated Compton-thick (CT; $N_{\mathrm{H}}>10^{24}\,\mathrm{cm^{-2}}$) one. Sources showing such a behaviour are sometimes referred to as X-ray ``changing look'' AGN \citep{2003MNRAS.342..422M}. These variations can be explained if at least part of the circumnuclear absorbing medium in type-2 AGN is clumpy \citep[][and references therein]{2002ApJ...571..234R,2010IAUS..267..299R,2012AdAst2012E..17B}, with the spectral changes caused by the transit of a cloud along the line of sight. 

In the standard unification picture, the obscuring torus should lie out of the LOS to the nucleus of type-1 AGN, so they are not expected to exhibit this behaviour. The heavily reddened type-1 AGN ESO 323-G77 has nonetheless displayed $N_{\mathrm{H}}$ variations from typical columns of $\sim 10^{23}\,\mathrm{cm^{-2}}$ to $1.5\times 10^{24}\,\mathrm{cm^{-2}}$ \citep{2014MNRAS.437.1776M}. The LOS in this source is suspected to graze the dusty torus, however, so that clumps therein could obscure the central engine, while leaving the broad line region visible. In addition, several cases of CT eclipses have been reported by \citet{2010IAUS..267..299R}. Furthermore, the archetypical type-1 AGN NGC 5548 shows strong evidence for fast outflows of ionized material that are likely launched from the accretion disk and can block large amounts of the soft X-ray emission \citep{2014Sci...345...64K}.

Type-1 AGN may also present themselves as Compton-thick in X-rays if there is a strong decrease of the nuclear X-ray luminosity, with a distant reflector, such as the pc-scale dusty torus itself, producing a light echo for some time \citep{2003MNRAS.342..422M}. This interpretation has been favoured in the case of the highly variable Seyfert 1 galaxy NGC 4051 \citep{1998MNRAS.301L...1G,1999MNRAS.307L...6U} and the Seyfert 1.9 galaxy NGC 2992 \citep{1996ApJ...458..160W,2000A&A...355..485G}. 

The current sample of AGN showing transitions between Compton-thin and -thick states is comprised of bright, local objects, for which high-S/N, multi-epoch X-ray observations have been obtained. At the higher redshifts probed by deep X-ray survey fields the situation is complicated by the fact that multiple observations with sufficient S/N are generally not available. Thus, even strong spectral variability of the kind discussed above may remain undetectable. We report here on the discovery of extreme spectral variations of this nature in an otherwise typical broad line type-1 QSO. In the following we assume a $\Lambda$CDM cosmology with $H_{0}=70\,\mathrm{km\,s^{-1} Mpc^{-1}}$, $\Omega_{\mathrm{m}}=0.3,$ and $\Omega_{\Lambda}=0.7$.

\section{Observations}
\label{sec:xobs}

The QSO RMID 278 ($\mathrm{RA}=214.32112$, $\mathrm{DEC}=+52.29764$) is part of the SDSS Reverberation Mapping (SDSS-RM) campaign \citep{2015ApJS..216....4S}. At $z=1.02$, it has a black hole mass of $M_{\mathrm{BH}}\sim 2.0\cdot 10^{8}M_{\sun}$ (based on single-epoch virial mass estimators), a bolometric luminosity of $L_{\mathrm{bol}}\sim 2.8\cdot 10^{45}\,\mathrm{erg\,s^{-1}}$ (estimated from $L_{3000}$ and mean bolometric correction from \citealt{2008ApJ...680..169S}), and an Eddington ratio of $\lambda_{\mathrm{Edd}}=L_{\rm bol}/L_{\rm Edd}=0.1$. 

RMID 278 was first covered in the X-rays by \XMM\, on 2000-07-20-21-23 with a total exposure of 80\,ks (ObsID 0127920401, 0127921001, and 0127921201; see \citealt{2001cghr.confE..65M}), with 247/178 (PN/MOS) net counts detected in the 0.5--8\,keV band. 
About five years later, the source region was covered by \Chandra\, with twelve observations, for a total of 189\,ks exposure time, taken between 2005-09-13 and 2005-12-11 as part of the \Chandra\, AEGIS-X deep survey of the Extended Groth Strip \citep{2009ApJS..180..102L}. The source was detected by \Chandra\, with 103 net counts in the 0.5--7\,keV band. 
 
\section{X-ray spectral analysis}
\label{sec:xspec}

We extracted X-ray spectra of RMID 278 from both XMM and \Chandra. The data reduction of the XMM observations and the extraction of X-ray source and background spectra for the PN/MOS detectors was performed following the methodology outlined in \citet{2011MNRAS.414..992G} and \citet{2016MNRAS.459.1602L}. The individual \Chandra\, spectra were extracted and merged into a single spectrum using the \texttt{ACIS EXTRACT} software package \citep{2010ApJ...714.1582B} following \citet{2014MNRAS.443.1999B}.

We analyzed the X-ray spectra using the Bayesian X-ray Analysis (BXA) software \citep{2014A&A...564A.125B}. The XMM PN/MOS spectra were fitted jointly in the observed-frame 0.5--8\,keV band and the EPIC PN/MOS background was modelled following \citet{2014A&A...561A..76M} \citep[see also][]{2016MNRAS.459.1602L}. The \Chandra\, spectrum was fitted in the observed-frame 0.5--7\,keV band, applying the background model of \citet{2014A&A...564A.125B}. Spectral models were absorbed by the average Galactic value of $N_{\mathrm{H,gal}}=1.11\times 10^{20}\,\mathrm{cm^{-2}}$ \citep{2005A&A...440..775K} in the direction of the source. Throughout this work, quoted fit parameters and error values correspond to the posterior median and standard deviation, energies are in the source rest-frame, and fluxes, luminosities and column densities are given in cgs units.   

\begin{figure*}
\centering

\subfloat{%
        \includegraphics[width=.48\textwidth]{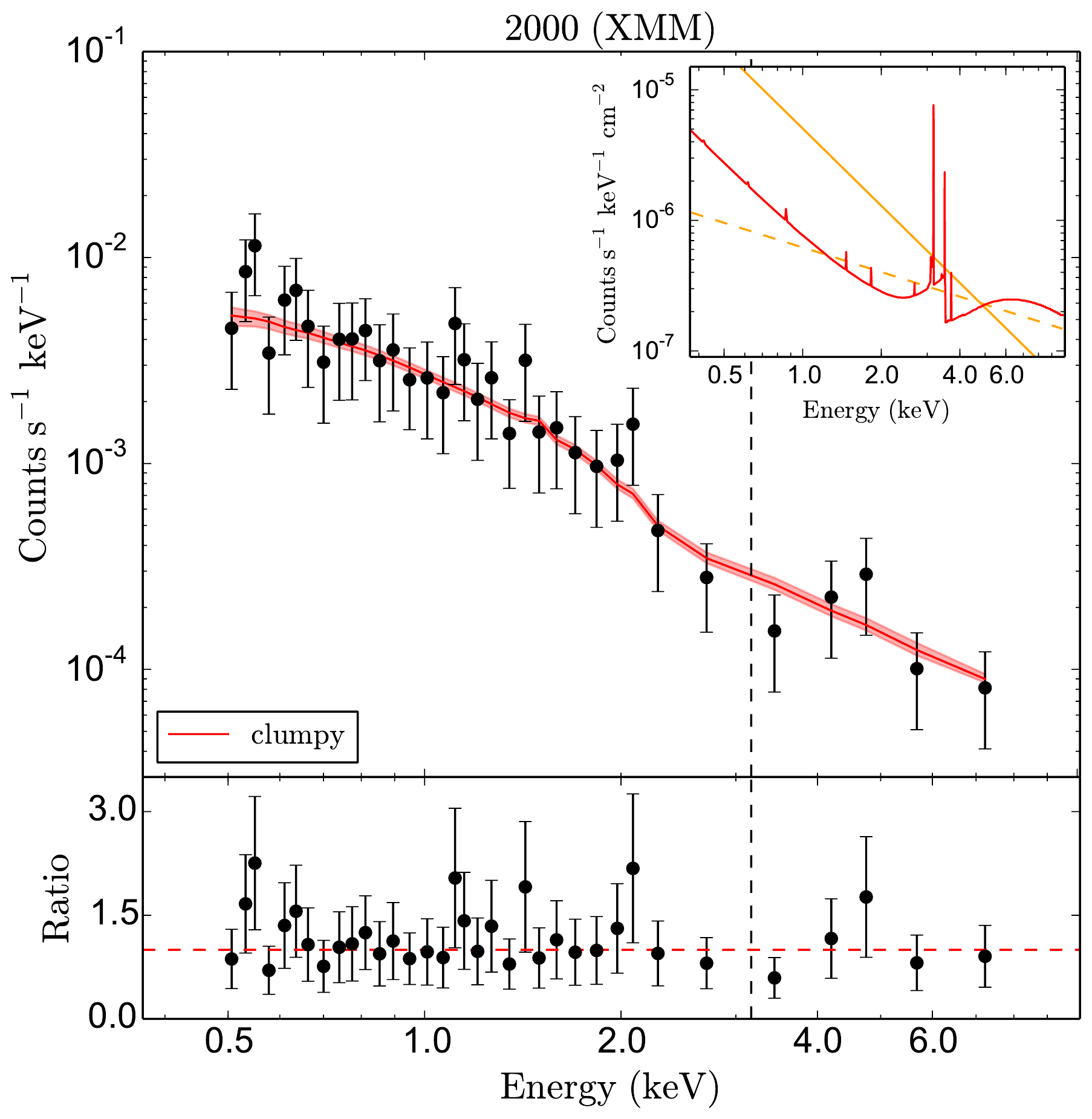}}
\hspace*{4mm}        
\quad
\subfloat{%
        \includegraphics[width=.48\textwidth]{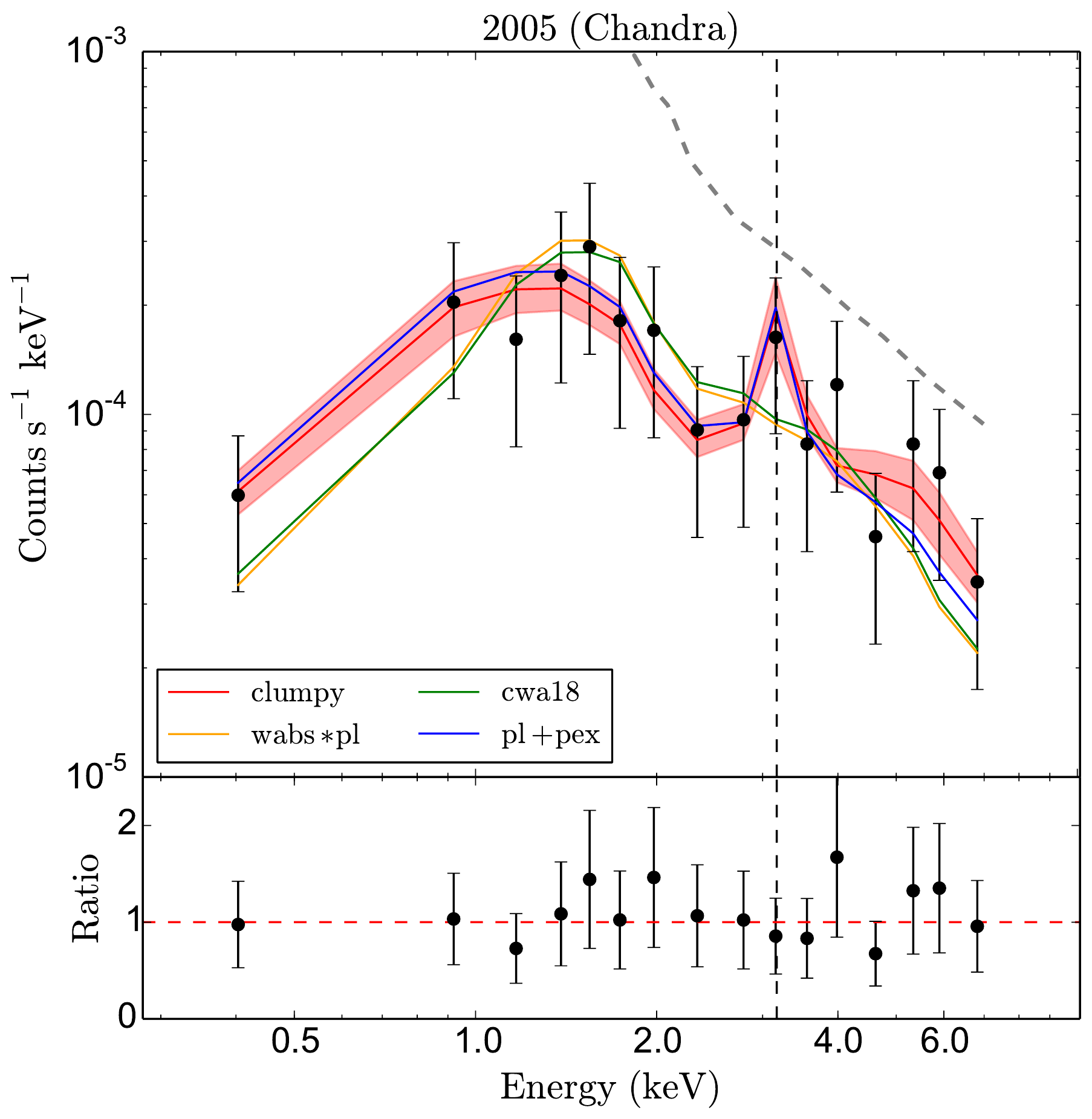}}
\caption{The dramatic X-ray spectral change of the type-1 QSO RMID 278. The XMM EPIC-PN spectrum from 2000 (left panel) is fully consistent with an unobscured power-law. In contrast, the \Chandra\, spectrum from 2005 (right panel) is extremely flat. The red curves in both panels display the best-fit \texttt{clumpy} model together with the $1\sigma$ error regions. The ratio of data and model is plotted in the lower panels, and the dashed vertical lines indicate the position of the Fe K$\alpha$ line at the source redshift of $z=1.02$. The insert in the left panel contains the unfolded spectral models fitted to the XMM spectrum (absorbed power-law with $\Gamma\sim 2$, orange curve) and the Chandra spectrum (absorbed power-law with $\Gamma\sim 0.6$, orange dashed curve; \texttt{clumpy} model, red curve). Three additional spectral models are shown in the right panel in different colors (see text for details), along with the \texttt{clumpy} fit to the XMM spectrum (dashed grey line). The observed spectra were grouped to 8 counts per bin for visualization only.}
\label{fig:specchange} 
\end{figure*} 

The unusual nature of RMID 278 was identified by comparing the photon indices of the subsample of 32 RM-QSOs having both XMM and \Chandra\, spectra available. The XMM spectrum from 2000 is well described by an unabsorbed power-law with $\Gamma=1.95\pm 0.09$ and small reflection fraction of $\log R=-1.20\pm 0.54$ (\texttt{pl$+$pex} model, see below), typical for a type-1 QSO, with source fluxes of $F_{\mathrm{0.5-2\,keV}}=(9.9\pm 1.0)\times 10^{-15}$ and $F_{\mathrm{2-10\,keV}}=(1.3\pm 0.1)\times 10^{-14}$\,$\mathrm{erg\,cm^{-2}\,s^{-1}}$, respectively. In contrast, the \Chandra\, spectrum from 2005 is extremely hard, with $\Gamma=0.62\pm 0.24$ and $\log R=-0.64\pm 0.76$ for an initial fit assuming a flat $\Gamma$-prior. Compared to the XMM observation, the soft band flux decreased by a factor of $7.6^{+3.3}_{-2.0}$ (0.5-2\,keV, $1\sigma$ min/max ranges), while the hard band flux decreased only by a factor of $2.5^{+0.5}_{-0.4}$ (2-10\,keV). The two spectra are shown in Fig. \ref{fig:specchange}, which illustrates the strong X-ray spectral variability of RMID 278.

To investigate different physical scenarios for the interpretation of the observed spectral variability, we tested various models for the \Chandra\, spectrum of RMID 278. We adopted a Gaussian $\Gamma$-prior with mean $\langle\Gamma\rangle=1.95$ and standard deviation $\sigma=0.15$, consistent with the observed distribution of the local AGN of \citet{1994MNRAS.268..405N}; for all other parameters of the models discussed below we assumed uniform or log-uniform priors. The Bayesian model comparison approach outlined in \citet{2014A&A...564A.125B} relies on the evidence $Z$ of a model: following the scale of \citet{jeffreys1961} we rule out a model $j$ with "strong evidence" if the evidence of the model differs by more than a Bayes factor of 10 from the total evidence of all compared models, corresponding to the condition $\log Z_{j}<\log (\sum_{i}Z_{i})-\log 10$. We compared the following five spectral models:
\begin{enumerate}
\item A power-law with no intrinsic absorption (\texttt{pl} model);
\item An absorbed power-law (\texttt{wabs$*$pl} model);
\item A power-law combined with a reflection component, modelled with \texttt{pexmon} \citep{2007MNRAS.382..194N}, i.e. \texttt{powerlaw}$+$\texttt{pexmon}, (\texttt{pl$+$pex} model). We define the reflection fraction, $R$, as the ratio between the normalization of the reflection component and that of the primary power-law . The inclination angle of the \texttt{pexmon} model was set to $60\degr$ and the photon index linked to that of the primary power-law. 
\item A warm absorber model \texttt{cwa18}, described in \citet{2007MNRAS.382..194N}, to test whether the spectral change can be better explained by the emergence of ionized absorption, potentially associated with outflows from the torus or disk \citep[see e.g.][]{2005A&A...431..111B}. The ionization parameter of this model was fitted in the interval $-1<\log\xi<3.5$.
\item The clumpy torus model (\texttt{clumpy}) of \citet{buchner_johannes_2018_1169181}, which self-consistently models Compton scattering, line fluorescence and absorption. This model was used to test for the possibility that the spectral change is caused by an eclipse of a dense cloud entering the LOS to the hot corona. The cloud distribution corresponds to a vertical Gaussian with standard deviation $\sigma$, consistent with existing CLUMPY \citep{2008ApJ...685..160N} infrared models. In addition, a covering parameter $C$ is included, describing a CT inner torus ring required to fit local CT AGN. Further parameters are the photon index of the primary power-law, the LOS $N_{\mathrm{H}}$ and a free viewing angle $\theta_{\mathrm{inc}}$. We also add a scattering component ($scat$) representing Thomson scattering of a volume-filling ionized medium, with a relative normalization to the primary power-law \citep{2006A&A...448..499B}.
\end{enumerate}

Apart from these five main spectral models, we tested and compared various other spectral models such as an absorbed power-law with a scattering but no reflection component (\texttt{wabs$*$pl+scatt}), the same model with a reflection component (\texttt{wabs$*$pl+pex+scatt}), the \texttt{BNtorus$+$pex$+$scatt} model of \citet{2014A&A...564A.125B}, and a warm absorber model with reflection and scattering components (\texttt{cwa18$+$pex$+$scatt})\footnote{By choosing the pexmon model we restricted ourselves to an optically thick reflection model, where reflection continuum hump and iron line emission are computed self-consistently. A yet more complex model, in which the line is produced in optically thin material, and thus essentially decoupled from the optically thick reflector, was not considered here for simplicity, given the low signal-to-noise of our spectrum.}. These models are discussed in less detail below. The results of the Bayesian model comparison of the five basic models and the fitted parameters of the two highest-evidence models are presented in Table~\ref{tab:modelpara}. We find that the models including a significant reflection component are clearly preferred, with the \texttt{pl} and \texttt{wabs$*$pl} models ruled out with high significance, whereas the \texttt{cwa18} model is not formally ruled out by our chosen threshold. The \texttt{clumpy} model has the highest evidence and is fitted with a LOS column density of $\log N_{\mathrm{H}}=24.31\pm 0.57$, suggesting a CT obscurer. In addition, the \texttt{clumpy} model yields a $\Gamma$ value that best agrees with the one derived from the XMM spectrum. However, the simpler \texttt{pl$+$pex} describes the \Chandra\, spectrum almost equally well, albeit with a slightly flatter photon index. Because of the limited number of counts in the \Chandra\, spectrum we could not significantly detect an Fe K$\alpha$ line; however, the $3\sigma$ upper limit for the equivalent width is $1$\,keV, still in agreement with expectations of $0.5-5$\,keV for a CT LOS \citep{2016ApJ...833..245B}. In Fig. \ref{fig:specchange} (right panel) the fitted models \texttt{wabs$*$pl}, \texttt{cwa18}, \texttt{pl$+$pex} and \texttt{clumpy} are plotted for comparison together with the \Chandra\, spectrum. 
\begin{table}
\caption{Fitted model parameters for the XMM and \Chandra\, spectra of RMID 278. For the former, the torus geometry and $scat$ parameters were fixed to the values derived from the latter. The model comparison results ($\Delta\log Z$ with respect to the \texttt{clumpy} model) for the \Chandra\, spectrum are shown at the bottom.}
\centering
\begin{tabular}{lcccc}
\hline
& \multicolumn{2}{c}{XMM} & \multicolumn{2}{c}{\Chandra}\\
Model & \texttt{pl$+$pex} & \texttt{clumpy} & \texttt{pl$+$pex} & \texttt{clumpy}\\
\hline
$\Gamma$ & $1.95\pm 0.09$ & $2.01\pm 0.09$ & $1.75\pm 0.12$  & $1.90\pm 0.14$ \\
$\log N_{\mathrm{H}}$ & - & $20.69\pm 0.38$ & - & $24.31\pm 0.57$ \\
$\log R$ & $-1.20\pm 0.54$ & - & $0.87\pm 0.13$  & - \\
$scat$ & - & $-1.29$ & - & $-1.29\pm 0.62$ \\
$\sigma$ & - & $31$ & - & $31\pm 20$ \\
${C}$ & - & $0.31$ & - & $0.31\pm 0.16$ \\
$\theta_{\mathrm{inc}}$ & - & $55$ & - & $55\pm 20$ \\
$\log L_{\mathrm{X}}$ & $43.83\pm 0.03$ & $43.84\pm 0.04$ & $43.13\pm 0.07$ & $44.90\pm 0.41$ \\
\hline
\end{tabular}
\begin{tabular}{lccccc}
Model & \texttt{clumpy} & \texttt{pl$+$pex} & \texttt{cwa18} & \texttt{wabs$*$pl} & \texttt{pl} \\
\hline
$\Delta\log Z$ & 0.00 & -0.34 & -0.83 & -1.41 & -1.59\\
ruled out & - & - & - & \checkmark & \checkmark \\
\hline
\end{tabular}
\label{tab:modelpara}
\end{table}
Adopting the \texttt{clumpy} model as our best model and fixing the torus geometry parameters to the values derived from the \Chandra\, spectrum, we find that the XMM spectrum is well-fitted by the model with $\Gamma=2.01\pm 0.09$ and $\log N_{\mathrm{H}}=20.69\pm 0.38$, consistent with an unobscured power-law. This result implies a dramatic column density variation by a factor of $>1000$, much larger than the $N_{\mathrm{H}}$ variability reported in previous works for both type-1 and type-2 AGN \citep[see e.g.][]{2002ApJ...571..234R,2014MNRAS.439.1403M}. Interestingly, comparing the absorption corrected 2-10\,keV power-law luminosities $\log L_{\mathrm{X}}=43.84\pm 0.04$ (XMM) and $\log L_{\mathrm{X}}=44.90\pm 0.41$ (\Chandra), suggests that despite the drop in observed flux the intrinsic luminosity actually increased in the \Chandra\, observation by a factor of $11.5^{+20.9}_{-7.4}$. 

Considering the second best model, \texttt{pl$+$pex}, the power-law 2-10\,keV luminosity decreased from $\log L_{\mathrm{X}}=43.83\pm 0.03$ (XMM) to $\log L_{\mathrm{X}}=43.13\pm 0.07$ (\Chandra), i.e., by a factor of $\sim 5.0^{+1.3}_{-1.0}$. This decrease could indicate a ``switched off'' corona. However, the large reflection flux (with $R=7.4^{+2.6}_{-1.9}$) in 2005, differs substantially from the earlier epoch, with the absolute flux of the reflection component increasing from $F_{\mathrm{2-10\,keV}}=1.1\times 10^{-16}$ (XMM, $1\sigma$ upper value $4.7\times 10^{-16}$) to $F_{\mathrm{2-10\,keV}}=2.8\times 10^{-15}$ (\Chandra, $1\sigma$ lower value $2.2\times 10^{-15}$) $\mathrm{erg\,cm^{-2}\,s^{-1}}$. We tested this explicitly by fitting the XMM spectrum with the \texttt{pexmon} parameters fixed to the same reflection flux as observed by \Chandra. This is ruled out by both the \texttt{pl$+$pex} fit with free \texttt{pexmon} parameters ($\Delta\log Z=-2.5$) and the \texttt{clumpy} fit to the XMM spectrum ($\Delta\log Z=-2.1$).

\begin{figure}
        \centering
        \includegraphics[width=.49\textwidth]{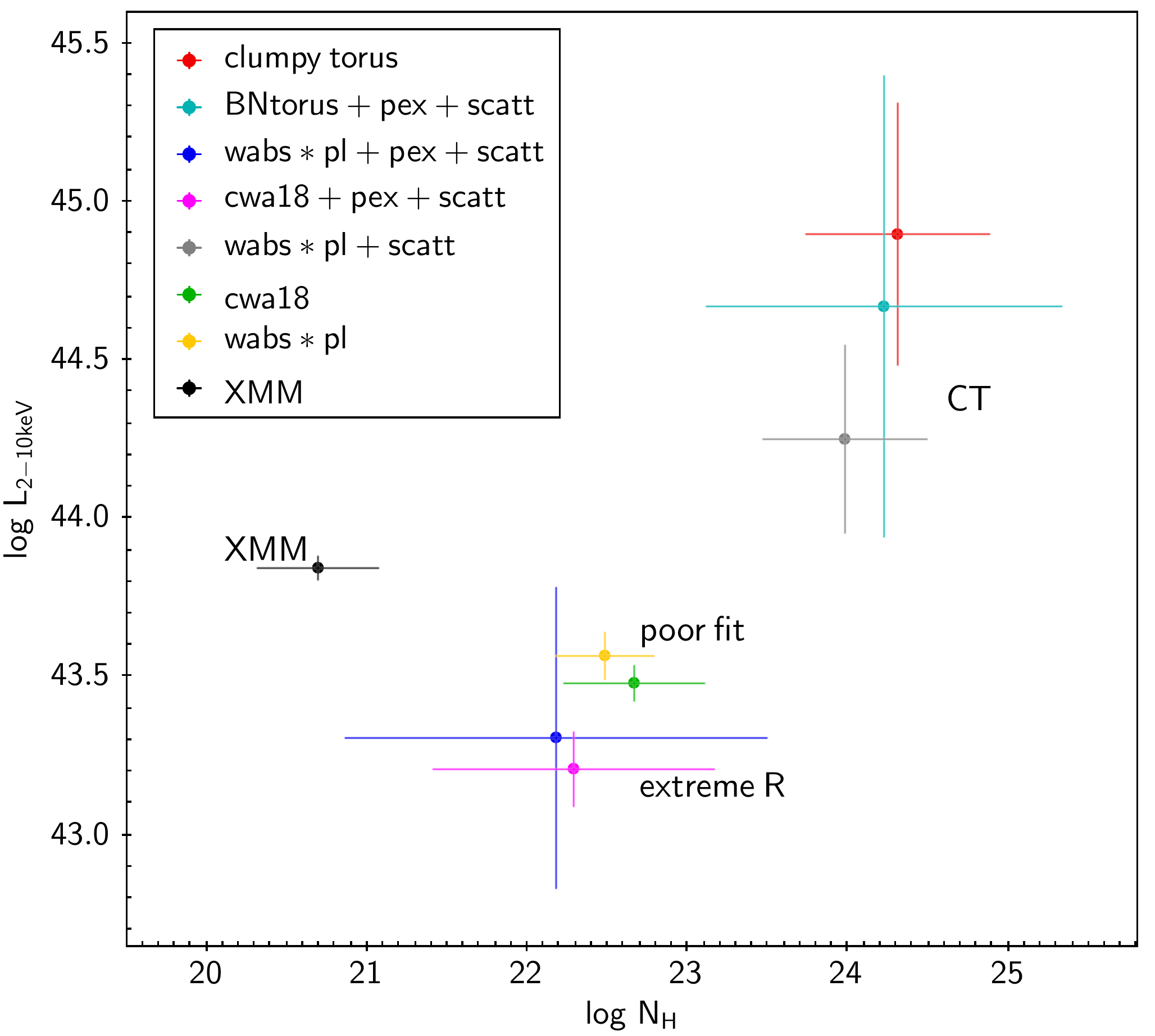}
        \caption{Intrinsic luminosity versus column density for various models (different colors) fitted to the \Chandra\, spectrum of RMID 278. The luminosity/column density pair of the \texttt{clumpy} model derived from the XMM spectrum is shown as black point for reference. Note that the blue and magenta points correspond to models with $\log R>0.5$. This provides evidence for a reflection-dominated state, irradiated with intrinsic luminosities of $\sim 10^{44} \mathrm{erg/s}$,
perhaps by a past corona that has now ``switched off''. The models of the yellow and green points have Bayes factors much lower than the \texttt{clumpy} model.}
        \label{fig:lumnh}
\end{figure}

Comparing the two aforementioned models with the resulting parameters from the \texttt{wabs$*$pl}, \texttt{cwa18}, \texttt{wabs$*$pl+scatt}, \texttt{wabs$*$pl+pex+scatt}, \texttt{BNtorus$+$pex$+$scatt}, and \texttt{cwa18$+$pex$+$scatt} models, we note some important similarities and differences. The fitted parameters of these additionally tested models are presented in Table ~\ref{tab:modelpara2}. First of all, none of these models can produce an intrinsic luminosity that is fully consistent with the value derived from the XMM spectrum. There are basically two groups of model solutions. Models with $\log N_{\mathrm{H}}<22.7$ and  $\log L_{\mathrm{X}}<43.6$ and models with $\log N_{\mathrm{H}}>24.0$ and $\log L_{\mathrm{X}}>44.2$. This is further demonstrated by Fig. \ref{fig:lumnh}, displaying the intrinsic luminosity/column density plane for all tested spectral models. The \texttt{wabs$*$pl+scatt} exhibits a CT column density and an $L_{\mathrm{X}}$ value that is larger than the XMM one, fully consistent within the errors with the \texttt{BNtorus$+$pex$+$scatt} and \texttt{clumpy} model parameters. The models including an ionized absorber have a poorly constrained ionization parameter and $L_{\mathrm{X}}$ values that are lower than the corresponding XMM value. Furthermore, the \texttt{cwa18$+$pex$+$scatt} model is fitted with a very low scattering normalization, a low intrinsic luminosity, and a very strong reflection component. In fact, the strong reflection normalization suppresses the ionized absorber part and the soft scattering component, which again provides evidence for a CT spectral state. We stress that any solution with $\log R>0$ requires extreme physical scenarios such as light bending, a ``switched off'' corona or a CT obscurer. Similarly, the \texttt{wabs$*$pl+pex+scatt} yields a low scattering normalization, a low intrinsic luminosity, a comparatively low column density but a very strong reflection component. The \texttt{BNtorus$+$pex$+$scatt} model, on the other hand, provides a fit solution that agrees very well with the self-consistent \texttt{clumpy} model. We point out that the Bayesian evidence values of the \texttt{wabs$*$pl+pex+scatt}, the \texttt{BNtorus$+$pex$+$scatt}, and the \texttt{clumpy} models agree within $\Delta\log Z=0.05$. Therefore, based on the Bayesian factors alone, one could not rule out any of these three models, but all three rule out the simple \texttt{wabs$*$pl} model and all three models require a strong reflection component and/or large column density increase compared to the XMM spectrum. The Bayesian evidence of the \texttt{cwa18$+$pex$+$scatt} model is $\Delta\log Z=0.15$ higher than for the \texttt{clumpy} model. In this case, the interpretation of the observations would require a significant decrease of the intrinsic luminosity (which is not observed in the optical light curve, see next section), accompanied by the emergence of an ionized absorber and a strong reflection component, which are both not detected in the XMM spectrum. The CT models instead suggest a strong column density and luminosity increase between the two epochs. Both interpretations are challenging to explain within a self-consistent physical scenario. 
\begin{table*}
\caption{Parameters fitted to the \Chandra\, spectrum of RMID 278 for various further spectral models.}
\centering
\begin{tabular}{lcccccc}
\hline
Parameter  & \texttt{wabs$*$pl} & \texttt{cwa18} & \texttt{wabs$*$pl+scatt} & \texttt{cwa18$+$pex$+$scatt} & \texttt{wabs$*$pl+pex+scatt} & \texttt{BNtorus$+$pex$+$scatt} \\
\hline
$\Gamma$ & $1.72\pm 0.26$ & $1.80\pm 0.14$ & $1.88\pm 0.14$ & $1.83\pm 0.15$ & $1.81\pm 0.13$ & $1.86\pm 0.14$ \\
$\log N_{\mathrm{H}}$ & $22.48\pm 0.30$ & $22.67\pm 0.44$ & $24.00\pm 0.51$ & $22.29\pm 0.88$ & $22.18\pm 1.32$ & $24.22\pm 1.11$ \\
$\log R$ & - & - & - & $0.65\pm 0.69$ & $0.62\pm 0.84$ & $0.05\pm 0.92$ \\
$\log\xi$ & - & $1.22\pm 1.23$ & - & $0.90\pm 1.14$ & - & - \\
$scat$ & - & - & $-1.17\pm 0.55$ & $-2.59\pm 1.19$ & $-2.28\pm 1.27$ & $-1.64\pm 0.97$ \\
$\log L_{\mathrm{X}}$ & $43.56\pm 0.08$ & $43.48\pm 0.06$ & $44.25\pm 0.30$ & $43.21\pm 0.12$ & $43.31\pm 0.48$ & $44.66\pm 0.73$ \\
\hline
\end{tabular}
\label{tab:modelpara2}
\end{table*}  
   
Summarizing our model comparison analysis, the X-ray data suggest that an eclipse by a dense absorber, or a ``switched off'' corona leaving a strong reflection signature, are the most likely scenarios to explain the dramatic spectral variability, though both of them require a large increase in the primary continuum flux prior to and/or during the Chandra observation.                       

\section{UV/optical properties of RMID 278}
\label{sec:uvoptprop}

\begin{figure}
        \centering
        \includegraphics[width=.49\textwidth]{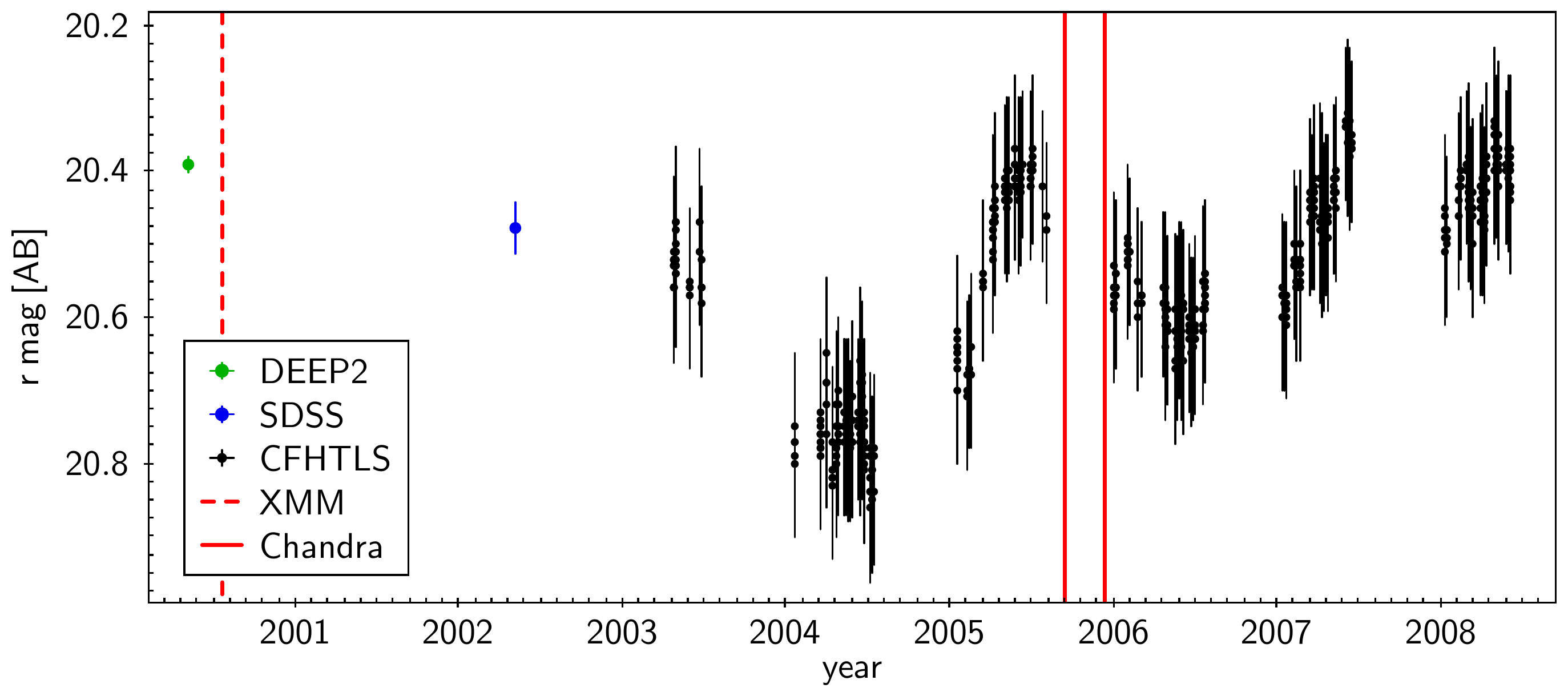}
        \caption{Optical (\textit{r} band) light curve combining observations from the DEEP2 (green), SDSS (blue), and CFHTLS-Deep surveys (black). The dates of the XMM and \Chandra\, (for first and last individual pointing) observations are marked by vertical lines.}
        \label{fig:rbandlc}
\end{figure}

To investigate the long-term variability of the source further, we collected optical photometry and spectroscopy from various observations closest in time to the X-ray observations. Fig. \ref{fig:rbandlc} shows the optical \textit{r} band (rest-frame $\lambda\sim 3100$\AA) light curve of RMID 278, with data taken from DEEP2 imaging \citep{2004ApJ...617..765C}, the SDSS--DR12 catalog \citep{2015ApJS..219...12A}, and the CFHTLS-Deep survey \citep[processed with the CFHTLenS pipeline; see][]{2013MNRAS.433.2545E}. The observed optical variability in the interval between the XMM and \Chandra\, observations corresponds to a maximum change in flux by a factor of $\sim 1.5$, consistent with typical QSO variability \citep[e.g.][]{2016A&A...585A.129S}. This is much less than the variability in optically selected changing look AGN  \citep[e.g.][]{2015ApJ...800..144L,2016MNRAS.457..389M}, and there is no hint of a large (factor $\sim 10$) continuum change implied by our X-ray spectral modelling. Of course, if the source underwent a heavily absorbed (Compton Thick) flare from the very innermost part of the disc/corona between 2000 and 2005, this might have been invisible in the optical band, whose light is generated by the outer accretion disc. We note that the intrinsic X/O ratio at the time of the XMM and \Chandra\, epochs is unknown, since optical data are only available several months before and after the X-ray observations. Nonetheless, assuming that the optical flux is similar to the flux observed closest to the X-ray observations, we obtain X/O(XMM)$=-0.2$ and X/O(\Chandra)$=-0.9$, using the 0.5-2\,keV flux and the i band optical flux. If we instead assume an intrinsic flux increase by a factor of 10 for the \Chandra\, spectrum, then X/O(Chandra)$=0.8$, again assuming the optical flux stayed the same. These values are all typical and within the expected locus for AGN of X/O$=\pm 1$ \citep[see e.g.][]{2012ApJS..201...30C}. Still, in the case of a CT obscurer, the actual values of the X/O ratios at both epochs depend strongly on the location of the obscurer, which is not determined from our data.  

\begin{figure}
        \centering
        \includegraphics[width=.48\textwidth]{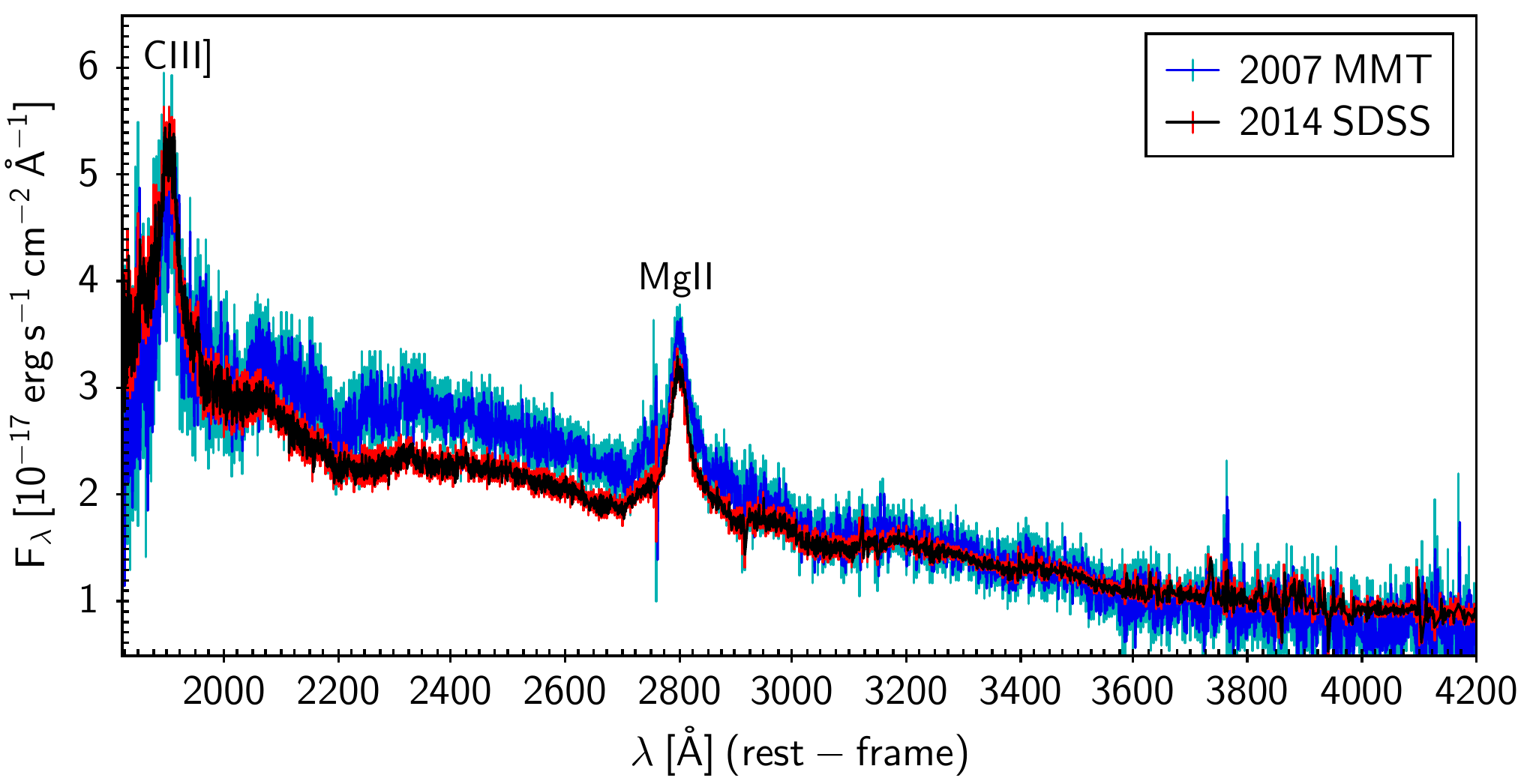}
        \caption{Comparison of the 2007 MMT (blue, $1\sigma$ errors in cyan) and the 2014 SDSS-RM spectrum (black, $1\sigma$ errors in red).}
        \label{fig:mmtvssdss}
\end{figure}

No optical spectra are available between 2000 and 2005, but the source was observed on 2007-03-16 with the MMT/Hectospec spectrograph \citep{2009ApJ...701.1484C} and has CFHT photometry with $r=20.45\pm 0.10$ measured just one day later. The MMT spectrum is shown together with the 2014 SDSS-RM spectrum (co-add of 32 spectra taken over a period of six month, \citet{2017AJ....154...28B}) in Fig. \ref{fig:mmtvssdss}. Both spectra display prominent broad MgII $\lambda\lambda\,2798$\AA\, and CIII] $\lambda\lambda\,1909$\AA\, emission lines and a strong blue continuum. No wind signatures, such as blueshifted absorption or reddening, are apparent.       

Thus, the optical light curve and the spectra disfavour a strong variation of the total accretion luminosity or any substantial optical obscuration as well as a change of the optical AGN type, unless these occurred during the periods not covered by the observations.  

\section{Discussion and conclusions}
\label{sec:discussion}

The repeat X-ray spectra of RMID 278 provide strong evidence for this typical $z\sim 1$ type-1 QSO switching from a transmission-dominated to a reflection-dominated state. This dramatic spectral change may be caused by either a nuclear eclipse event or a strong decrease of the corona luminosity, leaving the reflected emission from distant material as the dominant component of the source spectrum. In addition, the strong reflection component detected by \Chandra\, but not by XMM, requires a significant increase of the primary continuum X-ray flux in both scenarios, or a dramatic change in the source geometry. This is further supported by the relatively small decrease of the observed 2-10\,keV flux in between the two epochs.

In the ``switched off'' corona scenario, the X-ray reflection most likely comes from distant material, such as the pc-scale dusty torus. Its location can be estimated from the dust sublimation radius. Given the bolometric luminosity of our source, we obtain $R_{\mathrm{sub}}\sim 0.25-0.69$\,pc adopting either the relation fitted to observed K band time-lags \citep{2006ApJ...639...46S} or the theoretical prediction of \citet{1987ApJ...320..537B}. Based on these rough estimates, we would expect to still see a notable reflection echo in the 2005 spectrum, if the central engine switched off sometime between 2001 and 2005. On the other hand the strong increase in the reflection component between 2000 and 2005 would require a sustained and significant increase in the coronal luminosity prior to fading away, or a major change of the geometry of the reflector with respect to the corona in between the observations. The optical light curve (Fig.~\ref{fig:rbandlc}) displays no extreme variability between the X-ray observations, which, together with the optical spectra (Fig.~\ref{fig:mmtvssdss}), implies that the source probably did not undergo major changes in accretion rate or change its (optical) AGN type during this period. Even though AGN variability can be much more extreme in X-rays than in optical, especially on short timescales \citep{2014SSRv..183..453U}, these arguments tend to disfavour the ``switched off'' corona. 

The alternative hypothesis is that the source is in an X-ray bright, heavily obscured state at the time of the \Chandra\, observations. Within the best-fitting \texttt{clumpy} model interpretation, the event is most probably caused by a dense cloud eclipsing the X-ray corona, resulting in an $N_{\mathrm{H}}$ variation from the completely unobscured regime of $N_{\mathrm{H}}\sim 5\times 10^{20}\,\mathrm{cm^{-2}}$ to the CT regime with $N_{\mathrm{H}}\sim 2\times 10^{24}\,\mathrm{cm^{-2}}$. To the best of our knowledge, such an extreme $N_{\mathrm{H}}$ variation by more than three orders of magnitude has never before been observed in a type-1 QSO. This scenario also requires a large increase in the X-ray luminosity at the time of the \Chandra\, observation. Rather than being a coincidence, it could be that these events are connected. The increase in intrinsic luminosity implied by the X-ray spectrum could bring RMID 278 close to or perhaps even beyond the Eddington limit. Such a luminosity increase could launch a wind of CT clouds, which in turn obscures the hot corona \citep[see e.g.][]{1992ApJ...399L..23P}. However, the available data suggest that this would have been a dramatic and short (timescale few months) ``outburst'', possibly resulting in a ``failed wind'' that lifted CT material for only a short period of time. Due to the large uncertainties of the intrinsic luminosity in the CT solution, the implied luminosity increase should not be overinterpreted though. In any case, irrespective of the favoured interpretation of the observations, the data strongly suggest a dramatic change in the spectral properties of the source, associated to a significant change in the absorber column density, possibly accompanied by further changes in the source luminosity and/or relative geometry of the source/reflector system.

How often could these events occur? Based on a statistical analysis of X-ray eclipse events in local Seyfert galaxies monitored with RXTE, \citet{2014MNRAS.439.1403M} estimated the probability for a type-1 AGN to undergo an X-ray eclipse to be $0.6\%^{16.6\%}_{0.3\%}$. They did not detect any CT obscuration in their type-1 AGN sample (c.f. \citealt{2014MNRAS.437.1776M}) inferring an upper limit of $<15.8\%$ for such events. The CT obscuration in RMID 278 was seen for one out of 32 type-1 RM-QSOs with repeat XMM and \Chandra\, spectra. Given this limited data set, we are unable to derive robust constraints on the fraction of the type-1 QSOs undergoing CT eclipses. However, considering that we obtained spectra at just two epochs, and over a single time baseline of $\sim 5$ years, this behaviour could be very common. Comparing with the non-detections of \citet{2014MNRAS.439.1403M} and \citet{2014MNRAS.442.2116T} in nearby type-1 AGN, this could imply an important difference in the obscuration properties of local AGN and the more typical accreting SMBH found in deep X-ray surveys. Large area surveys with spectral timing information are needed to unveil the intrinsic fraction of CT obscuration events among the AGN population. The future surveys performed by SRG/eROSITA \citep{Merloni2012} will provide an unprecedentedly large sample of type-1 AGN with repeated X-ray observations on various timescales, allowing for a thorough statistical study of these extreme variations.

\section*{Acknowledgements}
JB acknowledges support from FONDECYT Postdoctorados grant 3160439 and the DFG cluster of excellence ``Origin and Structure of the Universe''. This research used the facilities of the Canadian Astronomy Data Centre operated by the National Research Council of Canada with the support of the Canadian Space Agency. 

%Funding for the SDSS III and IV has been provided by the Alfred P. Sloan Foundation, the U.S. Department of Energy Office of Science, and the Participating Institutions. SDSS acknowledges support and resources from the Center for High-Performance Computing at the University of Utah.

%%%%%%%%%%%%%%%%%%%%%%%%%%%%%%%%%%%%%%%%%%%%%%%%%%

%%%%%%%%%%%%%%%%%%%% REFERENCES %%%%%%%%%%%%%%%%%%

% The best way to enter references is to use BibTeX:

\bibliographystyle{mnras}
\bibliography{citations} % if your bibtex file is called example.bib

% Alternatively you could enter them by hand, like this:
% This method is tedious and prone to error if you have lots of references
%\begin{thebibliography}{99}
%\bibitem[\protect\citeauthoryear{Author}{2012}]{Author2012}
%Author A.~N., 2013, Journal of Improbable Astronomy, 1, 1
%\bibitem[\protect\citeauthoryear{Others}{2013}]{Others2013}
%Others S., 2012, Journal of Interesting Stuff, 17, 198
%\end{thebibliography}

%%%%%%%%%%%%%%%%%%%%%%%%%%%%%%%%%%%%%%%%%%%%%%%%%%

%%%%%%%%%%%%%%%%% APPENDICES %%%%%%%%%%%%%%%%%%%%%

%\appendix

%\section{Some extra material}

%If you want to present additional material which would interrupt the flow of the main paper,
%it can be placed in an Appendix which appears after the list of references.

%%%%%%%%%%%%%%%%%%%%%%%%%%%%%%%%%%%%%%%%%%%%%%%%%%

% Don't change these lines
\bsp	% typesetting comment
\label{lastpage}
\end{document}